\documentclass[12pt]{iopart}

\usepackage{iopams}
\usepackage{graphicx}
\begin{document}

\title{Unruh effect and Schwinger pair creation under extreme acceleration by ultraintense lasers}

\author{Chul Min Kim}
\address{Center for Relativistic Laser Science, Institute for Basic Science,
Gwangju 61005, Republic of Korea}
\address{Advanced Photonics Research Institute, Gwangju Institute of Science
and Technology, Gwangju 61005, Republic of Korea}
\ead{chulmin@gist.ac.kr}

\author{Sang Pyo Kim}
\address{Department of Physics, Kunsan National University, Kunsan 54150, Republic of Korea}
\address{Center for Relativistic Laser Science, Institute for Basic Science,
Gwangju 61005, Republic of Korea}
\ead{sangkim@kunsan.ac.kr}
\vspace{10pt}
\begin{indented}
\item[]November 15, 2017
\end{indented}

\begin{abstract}
A detector undergoing a huge acceleration measures a thermal distribution with the Unruh temperature out of the Minkowski vacuum. Though such huge accelerations occur naturally in astrophysics and gravity, one may design untraintense laser facility to detect the Unruh effect and simulate laboratory astrophysics. We derive the QED vacuum polarization and the vacuum persistence amplitude as well as the Schwinger pair creation in an accelerating frame when a constant electric field exists in the Minkowski spacetime. We advance a thermal interpretation of Schwinger pair creation in the Rindler space.
\end{abstract}

%
%
%
%
%

\section{Introduction}
A charged particle can accelerate by an electric field or gravitational field. In astrophysics, a massive particle undergoes a huge acceleration due to a gravitational source of compact star such as a neutron star or black hole. In terrestrial experiments, charges enormously accelerate by ultraintense lasers or by chromoelectric field temporarily generated during relativistic heavy ion collisions. Accelerations not only provide a mechanism for high energy particles but also induce a theoretical interest. One of the intriguing aspects of theoretical physics is the Unruh effect that is seen by a uniformly accelerating observer or detector for an infinite period of time \cite{Unruh:1976}. The accelerating observer measures a thermal spectrum out of the Minkowski vacuum since his or her worldline has a causally disconnected region, which forms part of a Cauchy surface in the Minkowski spacetime, and his or her vacuum becomes a mixed state after tracing out those states belonging to the causally disconnected region. The Unruh effect provides a way to understand the Hawking radiation from black holes \cite{Hawking:1975}. The Unruh temperature is proportional to the acceleration but the fundamental constants make the temperature extremely low except for a huge acceleration, $T_{\rm U} = a/(2.5 \times 10^{20})K$ in terms of Kelvin. Observing the Unruh effect thus requires a huge acceleration as summarized in Table 1.

\begin{table}[b]
	\caption{\label{tab1} Accelerations from lasers and gravity}
	\begin{tabular}{cc}
		\hline
		Linear acceleration & Gravitational acceleration \\
		\hline
		charge in an intense laser & particle at black hole horizon \\
		$a=\frac{c}{t_{\mathrm{pulse}}}$ & $\kappa=\frac{\left[c^{4}\right]}{4\left[G\right]M}$ \\
		femtosecond pulses & black holes \\
		$a=10^{23}~\mathrm{m/s^2}$ & $\kappa = 10^{13}\left( \frac{M_{S}}{M}\right) ~\mathrm{m/s^2}$ \\
		Unruh temperature $T_{\mathrm{U}}=\frac{\left[\hbar\right]a}{2\pi c\left[k_{\mathrm{B}}\right]}$ & Hawking temperature $T_{\mathrm{H}}=\frac{\left[\hbar\right]\kappa}{2\pi c\left[k_{\mathrm{B}}\right]}$ \\
		$T_{\mathrm{U}}=\frac{a}{2.5\times10^{20}~\mathrm{m/s^2}}\mathrm{K}$ & $T_{\mathrm{H}}=10^{-7}\left(\frac{M_{S}}{M}\right) \mathrm{K}$ \\
		\hline
	\end{tabular}
\end{table}

Another prominent prediction of quantum field theory is the spontaneous production of charged pairs known as the Schwinger mechanism in a strong electromgnetic field \cite{Schwinger:1951}. An intuitive understanding is that an electric field tilts otherwise an infinitely wide mass gap of the Diacr sea, allows a particle from the sea tunnels quantum mechanically through the barrier to a positive energy state and thus creates a pair of particle and antiparticle as a particle-hole pair. A cornucopia of Schwinger pairs are created when the electrostatic potential energy across one Compton wavelength is equal to the rest mass of charged particle, i.e. one pair per unit Compton volume for the critical field. In an electric field $E$ below the critical field $E_{\rm C}$, the mean number of created pairs is exponentially suppressed as ${\cal N} = e^{- \pi E_{\rm C}/E}$. A strong electromagnetic field makes the vacuum polarized due to the interaction of photons with the virtual pairs from the Dirac sea and leads to the Heisenberg-Euler and Schwinger effective action \cite{Schwinger:1951, Heisenberg:1936}, which is equivalent to summing over all one-loop diagrams. The physical predictions of the vacuum polarization in strong electromagnetic field are the vacuum birefringence and photon splitting, to name a few \cite{Piazza:2009}. In the in-out formalism initiated by Schwinger and DeWitt, the vacuum persistence amplitude (imaginary part of the effective action) is related to the mean number of Schwinger pairs since the vacuum instability due to Schwinger pair creation implies necessarily a complex effective action \cite{Kim:2008,Kim:2010}.

An accelerating charge will measure the intertwined effect both of the Unruh effect and the Schwinger mechanism. The real charge in a constant electric field experiences a uniform acceleration and feels a thermal state with the Unruh temperature while at the same time Schwinger pair creation occurs due to the electric field itself. In this paper, we study the Unruh effect and Schwinger pair creation under extreme acceleration and then propose a possible experiment using ultraintense lasers. The Schwinger mechanism of charged scalars was first studied in a two-dimensional spacetime by Gabriel and Spindel \cite{Gabriel:2000}, in which the vacuum persistence and mean number of charged pairs was explicitly calculated. We now advance a thermal interpretation of the vacuum persistence amplitude as well as the mean number in terms of the Unruh and Schwinger temperature.  An interesting theoretical model is an accelerating observer in a de Sitter space, in which the Unruh effect and the Gibbons-Hawking radiation are intertwined and lead to an effective temperature of the geometric mean of the Unruh and Hawking temperature \cite{Deser:1997}. The Schwinger effect was studied in a uniform electric field in (anti-)de Sitter space by Cai and Kim \cite{Cai:2014}, in which a thermal interpretation was first advanced.

\section{Extreme Acceleration by Ultraintense Lasers}

\begin{figure}
	\begin{center}
		\includegraphics[width=0.6\columnwidth]{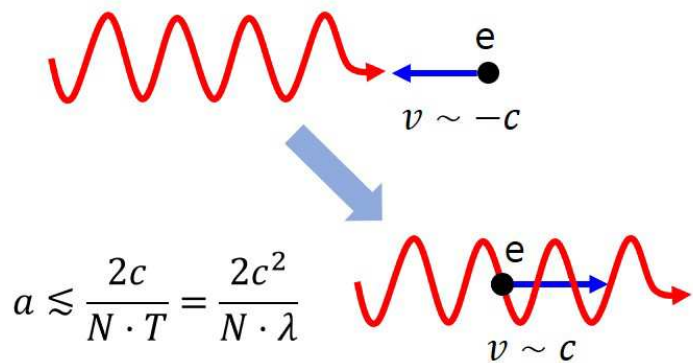}
	\end{center}
	\caption{ Average longitudinal acceleration in laser-electron collision. $T$ refers to the optical period, $\lambda$ to the laser wavelength, and $N$ to the number of the optical periods during which the electron changes its velocity from $-c$ to $c$. For $\lambda=800\ \mathrm{nm}$, $a \lesssim 2.3 \times 10^{22} g /N$. Depending on the laser intensity, $N$ can be smaller than 1.}
	\label{fig_la}
\end{figure}

Terrestrially, we usually accelerate charged particles with large-scale
accelerators, of which maximum electric field is limited by metallic
breakdown. The breakdown threshold is around 100 $\mathrm{MV/m}$,
and such strong field can induce an acceleration of $2\times10^{19}\: \mathrm{m/s^{2}}$
for nonrelativistic electrons; the corresponding Unruh temperature is below 0.1 K. Note that, for a given strength of
electric field, high-speed particles have lower acceleration than
low-speed particles due to relativistic mass increase. A much higher
field strength around 100 $\mathrm{GV/m}$ can be obtained from laser
wake field \cite{Esarey:2009}. However, such field can be exerted only on already
relativistic electrons comoving with the field, and thus acceleration,
defined as the change of velocity, cannot be sufficiently high due to the relativistic mass increase effect.

As a method to achieve sufficiently high acceleration for Unruh effect, we propose the collision
of a relativistic electron and an ultraintense laser pulse. When a
relativistic electron is reflected by an ultraintense pulse after
experiencing $N$ optical cycles of the pulse, changing its velocity from $-c$ to $ c$, the average acceleration
is around $2c/(NT)=2c^{2}/(N\lambda)$, where $c$ is the speed of
light, $T$ the optical period of the pulse, and $\lambda$ the wavelength
of the pulse. (See Fig. 1) For $\lambda=800\:\mathrm{nm}$, typical
for ultraintense lasers, the acceleration is $2\times10^{23}/N\:\mathrm{m/s^{2}}$,
and, with $N\sim O(1)$, it can lead to an Unruh temperature of around
1000 K. The corresponding Unruh radiation is in visible range and can be easily measured. As
a concrete example, we can consider the collision of a 1-MeV electron
(initial speed of 0.94$c$) and an ultraintense laser pulse with an intensity
of $2\times10^{20}\:\mathrm{W/cm^{2}}$. As the oscillation energy
of the electron under the laser pulse is 12.5 MeV, far larger than
the initial electron energy (1 MeV), $N$ would be an order of 1.
These parameters are easily available with current ultraintense lasers.

Although the proposal may look promising at a glance, it needs to be further analyzed in detail because it refers only to average longitudinal acceleration. The
actual electron motion is a superposition of longitudinal acceleration
and transverse oscillation, not a pure one-dimensional acceleration as considered in the simplest model of Unruh radiation. To use the proposed scheme, we need to develop a more sophisticated version of Unruh radiation in which Unruh radiation due to longitudinal motion and Larmor radiation due to transverse motion can occur simultaneousely. The differentiation of the two types of radiation should also be considered.

\section{Linear Uniform Acceleration and Unruh Effect}

There are four regions of the Minkowski spacetime classified by causality for the uniformly accelerating observer for an infinite period of time. The region accessible for the accelerating observer is the right Rindler (RR) wedge while the region causally disconnected from the right Rindler region is the left Rindler (LR) wedge. An electron accelerating in a uniform electric field in Minkowski spacetime follows a trajectory for an uniformly accelerating observer in the LR wedge while a positron under the same condition does follow a trajectory in the RR wedge. Note the LR and RR wedges form a Cauchy surface for the initial value problem in the Minkowski spacetime.

The Rindler metric for the RR and LR wedges is given by
\begin{eqnarray}
ds^2 = - (a \rho)^2 d \tau^2 + d \rho^2 + d^2 {\bf x}_{\perp},
\end{eqnarray}
where $a$ is the acceleration of the observer and ${\bf x}_{\perp} = (x,y)$. The Rindler coordinates in the RR wedge are
\begin{eqnarray}
t = \rho_{R} \sinh(a \tau), \quad z = \rho_{R} \cosh(a \tau), \quad (\rho_R \geq 0), \label{RR wedge}
\end{eqnarray}
and in the LR wedge are
\begin{eqnarray}
t = \rho_{L} \sinh(a \tau), \quad z = \rho_{L} \cosh(a \tau), \quad (\rho_R \leq 0).
\end{eqnarray}
The acceleration does not change the transverse plane, which means that the accelerating observer carries the transverse plane unchanged and the physics in higher dimensions is essentially the same as in two dimensions once the physical quantities corresponding to the transverse direction are separated.
The Killing vector $\partial/\partial t$ defines the Minkowski vacuum in the second quantization, while
the Rindler coordinates give another Killing vector $\partial/ \partial \tau$, which defines the Rinder vacuum.

The Bogoliubov transformations between the Fock spaces in the Rindler space and the Minkowski spacetime are
\begin{eqnarray}
\hat{a}^{\rm R}_{\omega {\bf k}_{\perp}} &=& \cosh \beta ~\hat{b}_{- \omega {\bf k}_{\perp}} + \sinh \beta~ \hat{b}^{\dagger}_{\omega -{\bf k}_{\perp}}, \nonumber\\ \hat{a}^{\rm L}_{\omega {\bf k}_{\perp}} &=& \cosh \beta ~\hat{b}_{ \omega {\bf k}_{\perp}} + \sinh \beta~ \hat{b}^{\dagger}_{-\omega -{\bf k}_{\perp}},
\end{eqnarray}
where
\begin{eqnarray}
\cosh \beta = \frac{e^{ \pi \omega/2a}}{2 \sinh(\pi \omega/a)}, \quad \sinh \beta = \frac{e^{ -\pi \omega/2a}}{2 \sinh(\pi \omega/a)}.
\end{eqnarray}
Then, the Minkowski vacuum defined with respect to the Killing vector $\partial/\partial t$ carries multi-particles of Rindler particles \cite{Crispino:2008}
\begin{eqnarray}
\langle 0_M \vert \hat{a}^{R \dagger}_{\omega {\bf k}_{\perp}} \hat{a}^{R}_{\omega' {\bf k}'_{\perp}} \vert 0_M \rangle =
\frac{1}{e^{\omega/T_{\rm U}} -1} \delta (\omega - \omega) \delta ({\bf k}_{\perp}- {\bf k}_{\perp}'),
\end{eqnarray}
where
\begin{eqnarray}
T_{\rm U} = \frac{[\hbar] a}{2 \pi [c] [k_{\rm B}]}
\end{eqnarray}
is the so-called Unruh temperature. The fundamental constants in square brackets will be set to unity hereafter.

\section{Scalar QED in Rindler Space}

Introducing $\rho_R = e^{a \xi}/a$, the metric for the RR wedge can be written in a conformal metric
\begin{eqnarray}
ds^2 = - dt^2 + dz^2 + d {\bf x}_{\perp}^2 = C^2(\xi) (- d \tau^2 + d \xi^2) + d {\bf x}_{\perp}^2.
\end{eqnarray}
where $C(\xi) = e^{a \xi}$. The two-dimensional metric for a uniform acceleration is conformal to the Minkowski spacetime. A constant electric field in the z-direction in the Minkowski spacetime, along which a positive charge accelerates, has the Maxwell tensor in a two-form
\begin{eqnarray}
{\cal F} = E dz \wedge dt = E e^{2a \xi} d \xi \wedge d \tau. \label{Maxwell}
\end{eqnarray}
The four-vector giving rise to the Maxwell tensor (\ref{Maxwell}) has the one-form
\begin{eqnarray}
A (\tau, \xi) = \frac{E}{2a} e^{2a \xi} d \tau. \label{gauge}
\end{eqnarray}

A complex scalar field minimally coupled to the Coulomb gauge (\ref{gauge}) obeys the Fourier-component equation
\begin{eqnarray}
\Bigl[\frac{d^2}{d\xi^2} + \Omega^2_{\omega, {\bf k}_{\perp}}  (\xi) \Bigr] \phi_{\omega, {\bf k}_{\perp}} (\xi) = 0, \label{field eq}
\end{eqnarray}
where
\begin{eqnarray}
 \Omega^2_{\omega, {\bf k}_{\perp}} = \Bigl(\frac{qE}{2a} e^{2a \xi} - \omega \Bigr)^2 - m^2 e^{2 a \xi}.
\end{eqnarray}
Here, $p_{\tau} = \omega$ $(\omega > 0)$ is a constant of motion. The equation (\ref{field eq}) describes the tunneling problem for Schwinger mechanism in the Coulomb gauge \cite{Kim:2002}. In the phase-integral method \cite{Kim:2007}, the leading terms for pair creation are obtained by summing over all independent contours of winding number one \cite{Kim:2014}
\begin{eqnarray}
N_{\omega, {\bf k}_{\perp}} = \Biggl| \sum_{J} e^{i \oint_{C_J} \Omega_{\omega, {\bf k}_{\perp}} (\xi) d\xi } \Biggr|.
\end{eqnarray}
Under a conformal mapping $\zeta = e^{2a \xi}$, a counterclockwise contour integral excluding a branch cut connecting two roots of the square root gives a leading term
\begin{eqnarray}
N_{\omega, {\bf k}_{\perp}}=  e^{-2 \pi \frac{\omega}{a}} e^{- \pi \frac{m^2 + {\bf k}^2_{\perp}}{qE}},
\end{eqnarray}
and another contour integral now enclosing a simple pole at $\zeta =0$ yields
\begin{eqnarray}
N_{\omega, {\bf k}_{\perp}}=  e^{- \pi \frac{m^2 + {\bf k}^2_{\perp}}{qE}}.
\end{eqnarray}

The in-vacuum and the out-vacuum in the RR wedge are related by the Bogoliubov transformation with the coefficients satisfying the relation
\begin{eqnarray}
|\alpha_{\omega, {\bf k}_{\perp}}|^2 - |\beta_{\omega, {\bf k}_{\perp}}|^2 =1,
\end{eqnarray}
and the mean number of spontaneously produced pairs is
\begin{eqnarray}
N_{\omega, {\bf k}_{\perp}} = |\beta_{\omega, {\bf k}_{\perp}}|^2.
\end{eqnarray}
The exact mean number follows from the second quantized field \cite{Gabriel:2000}
\begin{eqnarray}
N_{\omega, {\bf k}_{\perp}} = \frac{ e^{- 2 \delta} \bigl(1-  e^{-2 \gamma}  \bigr)}{1+  e^{-2 (\gamma + \delta)}},
\end{eqnarray}
where
\begin{eqnarray}
\gamma = \frac{\pi \omega}{a}, \quad \delta = \frac{\pi (m^2 + {\bf k}_{\perp}^2)}{2 qE}.
\end{eqnarray}

In the in-out formalism, the one-loop effective action for a scalar field is given by \cite{Kim:2008,Kim:2010}
\begin{eqnarray}
{\cal L}_{\rm eff} = i  \sum_{\bf K} \ln \bigl( \alpha^*_{\bf K} \bigr),
\end{eqnarray}
with ${\bf K}$ denoting all quantum states, while the vacuum persistence amplitude is
\begin{eqnarray}
2 {\rm Im} \bigl({\cal L}_{\rm eff} \bigr) = \sum_{\bf K} \ln \bigl( 1 + N_{\bf K} \bigr).
\end{eqnarray}
In fact, the vacuum persistence amplitude has an interesting form
\begin{eqnarray}
2 {\rm Im} \bigl({\cal L}_{\rm eff} \bigr) = \frac{qE}{2 \pi} \int \frac{d^2 {\bf k}_{\perp}}{(2\pi)^2} \Bigl[ \underbrace{ \ln \bigl(1 + e^{- 2 \delta} \bigr)}_{\rm scalar QED}  - \underbrace{\ln \bigl(1 + e^{ -2 (\gamma + \delta)} \bigr)}_{\rm QED + Unruh}  \Bigr]. \label{vac per}
\end{eqnarray}
In the zero-acceleration limit $(a=0, \gamma = \infty)$, Eq. (\ref{vac per}) reduces to the well-known Schwinger formula in a constant electric field. However, the zero-electric field limit does not recover the Unruh effect, which results from the Bogoliubov transformation between the Rindler vacuum and the Minkowski vacuum.
Using the reconstruction conjecture \cite{Kim:2017}, the vacuum polarization follows from the the vacuum persistence amplitude
\begin{eqnarray}
{\cal L}_{\rm eff} &=&  \frac{qE}{2 (2 \pi)} \int \frac{d^2 {\bf k}_{\perp}}{(2\pi)^2} \int_{0}^{\infty} {\cal P}
\int_0^{\infty} \frac{ds}{s} e^{-2 \delta s/\pi} \Bigl(1- e^{-2 \gamma s/\pi} \Bigr) \nonumber\\ && \times \Bigl(\frac{1}{\sin s} - \frac{1}{s} - \frac{s}{6} \Bigr).
\end{eqnarray}
Here, ${\cal P}$ denotes the principal value and the subtracted terms correspond to the renormalization of the vacuum energy and the charge.

In the two-dimensional Rindler space, there is no transverse momentum and the Schwinger effect is thus governed by the formula \cite{Gabriel:2000}
\begin{eqnarray}
N_{\omega} = \frac{e^{- \frac{m}{T_{\rm S} }} \Bigl(1 - e^{- \frac{\omega}{T_{\rm U}}} \Bigr) }{1+ e^{- \frac{m}{T_{\rm S}} } e^{- \frac{\omega}{T_{\rm U}} }}, \label{Sch Rin}
\end{eqnarray}
where
\begin{eqnarray}
T_{\rm S} = 2 \times \frac{qE/m}{2 \pi}, \quad T_{\rm U} = \frac{a}{2 \pi},
\end{eqnarray}
$T_{\rm S}$ being the effective temperature for the Schwinger effect in the Minkowski spacetime and $T_{\rm U}$ being the Unruh temperature. Note that the Schwinger effect (\ref{Sch Rin}) has applications in gravity. The near-horizon geometry of an extremal black hole in four dimensions has a product of a two-dimensional anti-de Sitter space and a two-sphere while the near-horizon geometry of a near-extremal black hole has an effect of a Rindler space as a measure of extremality. In fact, Eq. (\ref{Sch Rin}) appears in the Schwinger effect for a near-extremal charged black hole \cite{Kim:2016}.

\section{Conclusion}

We have studied the Schwinger pair creation and the vacuum polarization in a constant electric field in an accelerating frame described by a Rindler spacetime. The Schwinger mechanism for spontaneous production of pairs in an electromagnetic field has been known for long in the Minkowski spacetime and a constant electric field leads to the Boltzmann distribution of produced pairs with an effective temperature twice of the Unruh temperature for the acceleration of a charge by the electric field. An accelerating detector measures the Bose-Einstein or Fermi-Dirac distribution with the Unruh temperature from the Minkowski vacuum due to the presence of a causally disconnected region. The emission of charges in the Rindler spacetime thus reveals the Schwinger effect as well as the Unruh effect. We have proposed an experimental design for extreme accelerations using ultraintense lasers.

The Unruh effect is universal in curved spacetimes and has wide applications to astrophysical and gravitational phenomena. The Schwinger mechanism in a de Sitter space with a constant positive scalar curvature exhibits an intertwinement of the Schwinger pair creation and the Hawking radiation for the de Sitter space with an effective temperature determined by the geometric mean of the Unruh temperature for the charge acceleration and the Hawking temperature \cite{Cai:2014}. The Schwinger mechanism in an anti-de Sitter space is suppressed by the binding nature of the negative scalar curvature. The near-horizon geometry of charged black hole is approximated by a Rindler space with the acceleration determined by the surface gravity. Thus, the QED phenomena in the Rindler space have applications in astrophysics and gravity. The in-out formalism of spin 1/2-fermions and QED phenomena in the Rindler space will be addressed in a future publication \cite{Kim:2017b}.

\ack
The authors would like to thank Ehsan Bavasad and Cl\'{e}ment Stahl for useful discussions. This work was supported by IBS (Institute for Basic Science) under IBS-R012-D1, and by Gwangju Institute of Science and Technology under Research on Advanced Optical Science and Technology grant.

\
\

\end{document}